\documentclass[12pt]{iopart}

\usepackage{graphicx}
\usepackage{adjustbox}
\usepackage{aas_macros}

\usepackage{hyperref}
\usepackage{bm}
\usepackage{color}
\usepackage{amssymb}
\usepackage{breqn}

\usepackage{caption}
\usepackage{subcaption}

\def\aligned{\vcenter\bgroup\let\\\cr
\halign\bgroup&\hfil${}##{}$&${}##{}$\hfil\cr}
\def\endaligned{\crcr\egroup\egroup}

\begin{document}

\title[Joint search for isolated sources and a stochastic background in PTA data]{Joint search for isolated sources and an unresolved confusion background in pulsar timing array data}

\author{Bence B\'ecsy, Neil J. Cornish}

\address{eXtreme Gravity Institute, Department of Physics, Montana State University, Bozeman, Montana 59717, USA}
\ead{bencebecsy@montana.edu}
\vspace{10pt}
\begin{indented}
\item[]December 2019
\end{indented}

\begin{abstract}
Supermassive black hole binaries are the most promising source of gravitational-waves in the frequency band accessible to pulsar timing arrays. Most of these binaries will be too distant to detect individually, but together they will form an approximately stochastic background that can be detected by measuring the correlation pattern induced between pairs of pulsars. A small number of nearby and especially massive systems may stand out from this background and be detected individually. Analyses have previously been developed to search for stochastic signals and isolated signals separately. Here we present \texttt{BayesHopper}, an algorithm capable of jointly searching for both signal components simultaneously using trans-dimensional Bayesian inference. Our implementation uses the Reversible Jump Markov Chain Monte Carlo method for sampling the relevant parameter space with changing dimensionality. We have tested \texttt{BayesHopper} on various simulated datasets. We find that it gives results consistent with fixed-dimensional methods when tested on data with a  stochastic background or data with a single binary. For the full problem of analyzing a dataset with both a background and multiple black hole binaries, we find two kinds of interactions between the binary and background components. First, the background effectively increases the noise level, thus making individual binary signals less significant. Second, weak binary signals can be absorbed by the background model due to the natural parsimony of Bayesian inference. Because of its flexible model structure, we anticipate that \texttt{BayesHopper} will outperform existing approaches when applied to realistic data sets produced from population synthesis models.
\end{abstract}

%
%
%
%
%

\section{Introduction}
\label{sec:intro}

Pulsar Timing Arrays (PTAs) are designed to detect gravitational waves (GWs) in the nanohertz regime by looking for signatures of a GW signal in the pulse arrival times of regularly observed millisecond pulsars (see e.g. \cite{pta_review}). Currently there are three major PTA projects: the North American Nanohertz Observatory for Gravitational Waves (NANOGrav, \cite{nanograv11data}), the European Pulsar Timing Array (EPTA, \cite{epta_data}), and the Parkes Pulsar Timing Array (PPTA, \cite{ppta}). These groups collaborate under the aegis of the International Pulsar Timing Array (IPTA, \cite{ipta, ipta_dr1}).

The dominant GW sources in the frequency band of PTAs are expected to be supermassive black-hole binaries (SMBHBs). The collection of these SMBHBs is expected to give rise to a stochastic gravitational wave background (GWB), which has been searched for in previous PTA datasets (see e.g.~\cite{nanograv11gwb, epta_gwb}). The scarcity of SMBHBs in the local Universe can introduce anisotropy in the GWB~\cite{Sesana:2008mz,2011MNRAS.411.1467K,Ravi:2012bz,local_nhz_landscape}, which makes the standard isotropic searches sub-optimal, thus requiring changes to those standard algorithms \cite{Cornish:2013aba, anisotropy_method, anisotropy_epta}.

In addition, some of the brightest sources might be individually detected as so called continuous waves (CWs), referring to the fact that the frequency of GWs from an SMBHB does not change appreciably during the observation time of PTAs. While it is more likely that PTAs will detect the GWB before detecting any CW signals \cite{nature_of_first_det}, searches have been carried out resulting in upper limits on the possible amplitude of CWs (see e.g.~\cite{nanograv11cw, epta_cw}).

In the context of the Laser Interferometer Space Antenna (LISA, \cite{lisa}), it has been shown that removing the brightest sources results in a Gaussian background from the unresolved sources~\cite{Timpano:2005gm}. For PTAs, it has been shown that removing the brightest CW signal significantly reduces the anisotropy of the GWB~\cite{local_nhz_landscape}. Describing a signal as either a collection of individual components or as a stochastic background has been considered in the context of a toy model using a pair of co-located LIGO \cite{aLIGO} detectors and a signal formed from multiple sinusoids~\cite{when_stoch}. The study considered three different models: i) a collection of sinusoidal GWs; ii) a Gaussian stochastic GWB; iii) the combination of i) and ii). It was found that model iii) was almost always preferred, despite the fact that the dataset was generated with the assumption of model i). This is because of the natural parsimony of Bayesian statistics preferring a model with fewer parameters even if it fits the data slightly worse. Non-Bayesian methods have been proposed to search for multiple CWs in PTA data in \cite{multi_cw1, multi_cw2}.

These previous studies motivate a joint search for CWs and an isotropic GWB. Here we present a method for such a search using Bayesian inference and a trans-dimensional Reversible Jump Markov Chain Monte Carlo (RJMCMC) sampler~\cite{rjmcmc}. RJMCMC methods are used in other GW data analysis problems as well, like the detection of unmodeled transient GWs in LIGO data by the BayesWave algorithm \cite{bayeswave}, and the detection of galactic binaries in LISA data \cite{Umstatter:2005, Crowder:2006eu,Littenberg:2011zg, Littenberg:2020}. Such a sampler is able to explore a parameter space of varying dimension, allowing it to decide what to include in the model based on the data. In our case it can turn the GWB component on or off, and can also change how many CWs are in the model (including the possibility of zero CWs).

The paper is organized as follows: Section \ref{sec:methods} introduces the models we use to describe our PTA dataset and how our RJMCMC sampler explores the whole parameter space including different model dimensions. Section \ref{sec:results} presents the results of analyzing various simulated datasets, which we carried out to assess the performance of our algorithm. We provide our conclusions and discuss future work in Section \ref{sec:conclusion}.

\section{Methods}
\label{sec:methods}

In this section, we discuss our signal and noise model, and introduce the RJMCMC sampler we developed to explore the variable-dimension parameter space in question.

\subsection{Models}
\label{sec:model}

We use the \texttt{enterprise}\footnote{\url{https://github.com/nanograv/enterprise}} software package for handling the pulse time-of-arrival (TOA) data and for calculating the likelihood for our Bayesian analysis. Timing models were produced by \texttt{libstempo}\footnote{\url{https://github.com/vallis/libstempo}}, which is a \texttt{python} wrapper for the \texttt{tempo2}\footnote{\url{https://bitbucket.org/psrsoft/tempo2/}} timing package. After applying the timing model, we get a set of timing residuals for each pulsar. We model the $i$th residual in the $k$th pulsar as:
\begin{equation}
 \delta t_{ki} = \sum_l M_{kil} \delta \xi_{kl} + n_{ki} + g_{ki} + s_{ki}(\bm{\theta}),
 \label{eq:model}
\end{equation}
where $\delta \xi_{kl}$ is the offset between the true timing model parameters and their best fit parameters, and $M_{kil}$ is the design matrix, which represents a timing model linearized around the best fit parameters. $n_{ki}$ is the noise present in the TOAs, $g_{ki}$ is the contribution of the isotropic GWB, while $s_{ki}(\bm{\theta})$ describes the response to CWs, where $\bm{\theta}$ is the set of parameters describing the CWs. From these four components, three of them are modeled directly (see following subsections), while the timing model component is analytically marginalized over~\cite{vanHaasteren:2008yh}. In the following sections we supress the $i$ index to simplify the notation.
\subsubsection{Noise model}

The noise in the TOAs is assumed to be Gaussian, uncorrelated between pulsars, and it can be decomposed in the $k$th pulsar as:
\begin{equation}
 n_k = n^{\rm WN}_k + n^{\rm RN}_k,
 \label{eq:noise}
\end{equation}
where $n_k^{\rm WN}$ is the white noise (WN) and $n_k^{\rm RN}$ is the red noise (RN) in the $k$th pulsar. We model the covariance matrix of $n_k^{\rm WN}$ as:
\begin{equation}
 C^{\rm WN}_{k,ij} = E_k^2 W_{ij},
 \label{eq:wn_cov}
\end{equation}
where $E_k$ is the ``EFAC'' parameter from \texttt{TEMPO2}, and $W = {\rm diag}\ {\sigma_i^2}$, where $\sigma_i$ is the nominal uncertainty of the $i$th TOA. We used a uniform prior on each ``EFAC'' value in the range of $[0.01, 10]$. This is a simplified version of the WN model commonly used in PTA data analysis (see e.g.~\cite{nanograv9gwb}), which can be extended easily if we want to analyze real PTA data. 

To model $n_k^{\rm RN}$, we follow the procedure commonly used by NANOGrav (see e.g.~\cite{nanograv11cw}). We describe the RN power spectral density (PSD) with a power-law model:
\begin{equation}
 P^{\rm RN}_k (f) = \left(A^{\rm{RN}}_k\right)^2 \left( \frac{f}{f_{\rm{yr}}} \right)^{- \gamma_k},
 \label{eq:rn_psd}
\end{equation}
where $A^{\rm{RN}}_k$ is the RN amplitude, and $\gamma_k$ is the spectral index in the $k$th pulsar, while $f_{\rm{yr}} \equiv 1/(1 \rm{yr})$. We use 30 frequency bins spaced linearly between $1/T_{\rm{obs}}$ and $30/T_{\rm{obs}}$, where $T_{\rm{obs}}$ is the total observation time. Note that $30/T_{\rm{obs}}$ is smaller than our Nyquist frequency, but we only need to model RN at low frequencies, because WN will dominate at higher frequencies. Since the simulated datasets used in this study (see Section \ref{sec:results}) does not include RN, including it in the model is only motivated by the fact that it is similar to the GWB model (the difference being in the correlations, see Section \ref{sec:gwb}), so it will provide a competing model. This motivates us to simplify the RN model by requiring it to be a common red process for all pulsars, i.e.~$A^{\rm{RN}}_k = A^{\rm{RN}}$ and $\gamma_k=\gamma$ for all $k$. This assumption will simplify our modeling, and speed up sampling, but can be easily relaxed in the future when analyzing actual PTA data. We used a uniform prior on the amplitude $A^{\rm RN} \in [10^{-18}, 10^{-13}]$ and the spectral index $\gamma \in [0,7]$.

\subsubsection{GWB model}
\label{sec:gwb}

The timing residuals resulting from a GWB are described by a Gaussian process, which is correlated between pulsars, and described by the following cross-power spectral density:
\begin{equation}
 S^{\rm{GWB}}_{kl}(f) = \Gamma_{kl} \frac{\left(A^{\rm GWB}\right)^2}{12 \pi^2} \left( \frac{f}{f_{\rm{yr}}} \right)^{- \gamma},
\end{equation}
where $\Gamma_{kl}$ is the overlap reduction function characterizing the correlations between the $k$th and $l$th pulsars. In our case, where we consider an isotropic GWB produced by SMBHBs, the overlap reduction function is given by the Hellings and Downs curve \cite{hd_corr}. $A^{\rm GWB}$ is the amplitude of the GWB, and $\gamma=13/3$ is the expected spectral index of the GWB \cite{gwb_spec_idx, Sesana:2008}, which we keep fixed in our analysis. However, our analysis can be easily extended to vary $\gamma$. We use the same 30 frequency bins as for the RN. We used a uniform prior on the amplitude $A^{\rm GWB} \in [10^{-18}, 10^{-13}]$.

\subsubsection{CW model}
\label{sec:cw}
In the $k$th pulsar, we model multiple CW signals as:
\begin{dmath}
 s_k(t; \bm{\theta})= \sum_{m=1}^{N_{\rm{CW}}} \left[ F^{+}(\bm{\theta}_m, \theta_k, \phi_k) \Delta s_k^{+} (t; \bm{\theta}_m) + F^{\times}(\bm{\theta}_m, \theta_k, \phi_k) \Delta s_k^{\times} (t; \bm{\theta}_m)\right],
\end{dmath}
where $N_{\rm{CW}}$ is the number of CW sources included in the model, $\bm{\theta}_m$ are the parameters describing the $m$th CW source, $F^{+,\times} (\bm{\theta}_m, \theta_k, \phi_k) = F^{+,\times} (\theta_m^{\rm{GW}}, \phi_m^{\rm{GW}}, \theta_k, \phi_k)$ are the antenna pattern functions given e.g.~by Eq.~(7) and (8) in \cite{nanograv11cw}, where $\theta_m^{\rm{GW}}$ and $\phi_m^{\rm{GW}}$ are spherical coordinates of the $m$th CW source, while $\theta_k$ and $\phi_k$ are spherical coordinates of the $k$th pulsar. $\Delta s_k^{+,\times}$ is the difference between the metric perturbation at the Earth (``Earth term'') and at the $k$th pulsar (``pulsar term''). Due to different pulsar distances, the pulsar terms do not add up as coherently as the Earth terms, so for simplicity, we only include the Earth terms in this study. Pulsar terms can be easily included in our analysis, with the expense of a few additional parameters we need to sample. This simplifications makes $\Delta s_k^{+,\times} (t; \bm{\theta}_m) = -s^{+,\times}(t; A_{\rm CW}, \omega^{\rm{GW}}, \iota, \psi, \Phi_0)$, where $s^{+,\times}$ are the $+$ and $\times$ polarization Earth terms given by:
\begin{dmath}
 s^{+} (t) = \frac{A_{\rm CW}}{\omega^{\rm{GW}}} \left[ - \sin (\omega^{\rm{GW}} t + 2 \Phi_0) \left( 1+\cos^2 \iota \right) \cos 2 \psi - 2 \cos (\omega^{\rm{GW}} t + 2 \Phi_0) \cos \iota \sin 2 \psi \right],
\end{dmath}

\begin{dmath}
 s^{\times} (t) = \frac{A_{\rm CW}}{\omega^{\rm{GW}}} \left[ -\sin (\omega^{\rm{GW}} t + 2 \Phi_0) \left( 1+\cos^2 \iota \right) \sin 2 \psi + 2 \cos (\omega^{\rm{GW}} t + 2 \Phi_0) \cos \iota \cos 2 \psi \right],
\end{dmath}
where $A_{\rm CW}$ is the GW amplitude, $\omega^{\rm{GW}} = 2 \pi f^{\rm{GW}}$ is the GW angular frequency, $\iota$ is the inclination angle, and $\psi$ is the polarization angle, and $\Phi_0$ is the initial orbital phase of a CW source. We can see that our CW model is described by 7 $N_{\rm{CW}}$ parameters, and the parameters of a single source are $\bm{\theta}_m = \{ A_{{\rm CW},m}, f^{\rm{GW}}_m, \iota_m, \psi_m, \Phi_{0, m}, \theta_m^{\rm{GW}}, \phi_m^{\rm{GW}} \}$. We used a uniform prior on the amplitude $A_{\rm CW} \in [10^{-18}, 10^{-13}]$, the frequncy $f^{\rm{GW}} \in [3.5 \ {\rm nHz}, 100 \ {\rm nHz}]$, the polarization angle $\psi \in [0 , \pi]$, the initial phase $\Phi_0 \in [0 , 2 \pi]$, and the cosine of the inclination angle $\cos \iota \in [-1 , 1]$. We also used a uniform on the sky prior resulting in a uniform prior on $\phi^{\rm{GW}} \in [0 , 2 \pi]$ and on $\cos \theta^{\rm{GW}} \in [-1 , 1]$.

\subsection{RJMCMC}

As described in Section \ref{sec:model}, the number of parameters in our model is:
\begin{equation}
 N (N_{\rm{CW}}, N_{\rm{GWB}}) = N_{\rm{RN}} + N_{\rm{psr}} + 7 N_{\rm{CW}} + N_{\rm{GWB}},
\end{equation}
where $N_{\rm{RN}}=2$ is the number of parameters used to describe RN, $N_{\rm{psr}}$ is the number of pulsars in the array, which corresponds to the number of parameters used in the WN model, $N_{\rm{CW}}\in [0, N^{\rm{max}}_{\rm{CW}}]$ is the number of CWs included in the model, where $N^{\rm{max}}_{\rm{CW}}$ is the maximum number of CWs allowed, and $N_{\rm{GWB}}=\{0,1\}$ indicates if a GWB is included or not. The GWB is described by only one parameter, $A^{\rm GWB}$, because we fix $\gamma=13/3$.  Note that $N_{\rm{CW}}$ and $N_{\rm{GWB}}$ are allowed to vary, so the dimensionality of our parameter space is not fixed. We use a uniform prior on both $N_{\rm{CW}}$ and $N_{\rm{GWB}}$.

To sample this variable-dimension parameter space we implemented an RJMCMC (also called trans-dimensional MCMC, see e.g.~\cite{rjmcmc, rjmcmc_new, rjmcmc_thesis, rjmcmc_example}) sampler called \texttt{BayesHopper}\footnote{\url{https://github.com/bencebecsy/BayesHopper}}, which treats $N_{\rm{CW}}$ and $N_{\rm{GWB}}$ as free parameters allowing to sample the full parameter space. In the rest of this section, we describe different components of this algorithm.

Note that a similar data analysis problem arises in the context of LISA, where one needs to model a collection of signals from tens of thousands of galactic binaries (see e.g.~\cite{Umstatter:2005, Crowder:2006eu, Littenberg:2011zg, Littenberg:2020}). As a result, the sampling techniques we use show many similarities to those used in \cite{Littenberg:2020}.

\subsubsection{Fisher matrix proposals}
To achieve good mixing within a given model, we use a proposal based on the Fisher information matrix, defined as:
\begin{equation}
 \Gamma_{ij} = -\partial_i \partial_j \ln p(d|\bm{\theta}), 
\end{equation}
where $p(d|\bm{\theta})$ is the likelihood, and  $\partial_i$ indicates partial derivative with respect to $\theta_i$. We calculate $\Gamma_{ij}$ every few hundred iterations numerically, since it is changing as we move around in the parameter space. If $p(d|\bm{\theta})$ is exactly Gaussian, $\Gamma_{ij}$ gives the inverse of the covariance matrix, but the proposal works well even for non-Gaussian distributions.

When attempting a Fisher matrix based step, the proposed $(i+1)$th sample is given by:
\begin{equation}
 \bm{x}_{i+1} = \bm{x}_{i} + \xi \bm{e}_j,
\end{equation}
where $\bm{e}_j$ is the $j$th normalized eigenvector of $\Gamma_{ij}^{-1}$, and $\xi \sim \mathcal{N}(0,\Delta_j^2)$, where $\Delta_j$ is the $j$th eigenvalue of $\Gamma_{ij}^{-1}$. We choose $j$ randomly for each proposed step. If the target distribution is not too different from a multivariate Gaussian, this proposal is practically a draw from the marginal distribution along a principal axis, thus we get optimal acceptance and mixing of the chain. We find that in this application, the posterior distributions around the maximum are close enough to being Gaussian that this proposal results in very good mixing.

\subsubsection{Global proposal using the $\mathcal{F}_e$-statistic}
\label{ssec:Fe}
The CW model has 7 parameters, some of which are very well constrained. This means that starting from a random point in the parameter space, it takes many iterations to find the maximum of the posterior using our Fisher matrix based proposal (or any local proposal). This is especially problematic in an RJMCMC such as this, because the sampler might only spend a few hundred samples in one model before changing the number of parameters, so the sampler should converge quickly.

\begin{figure}[htb]
 \centering
   \includegraphics[width=1.0
   \textwidth]{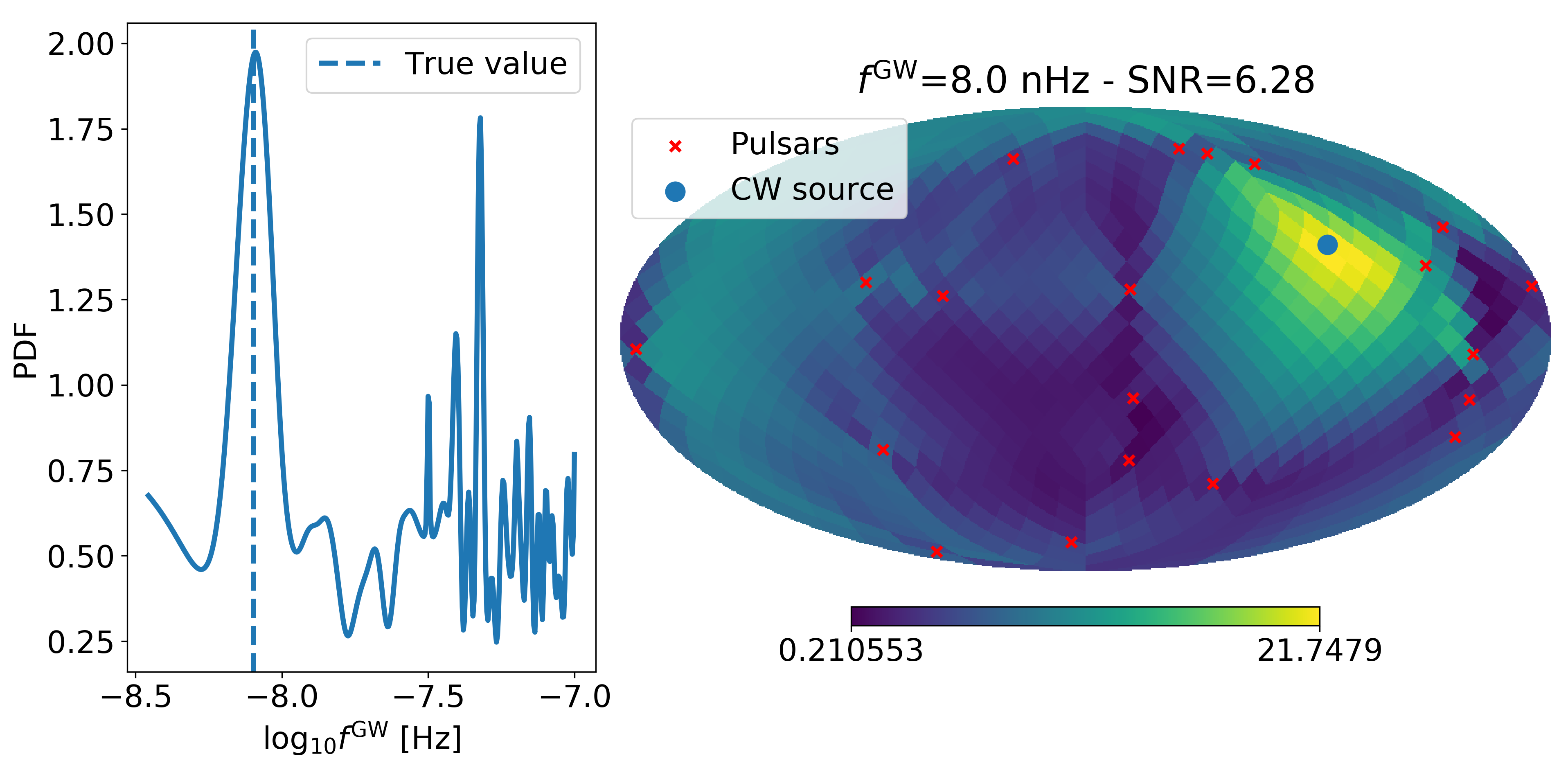}
 \caption{Global proposal for the dataset with an $A_{\rm CW} = 7.08 \times 10^{-15}$ simulated CW signal used in Section \ref{ssec:study2}. \emph{Left:} $\mathcal{F}_e$ marginalized over sky location showing the PDF of proposing a given value of $\log_{10} f^{\rm GW}$. \emph{Right:} $\mathcal{F}_e$ at the true frequency. Positions of the simulated source and pulsars in the array are indicated by the blue circle and red crosses, respectively. Note how strongly $\mathcal{F}_e$ concentrates around the true sky location and true frequency of the CW source, indicating that it is much more efficient as a proposal than just drawing sky location and frequency from a uniform distribution.}
 \label{fig:f-stat_map}
\end{figure}

To help with convergence we introduce a global proposal based on the Earth-term $\mathcal{F}$-statistic ($\mathcal{F}_e$, see \cite{pta_fstat}). $\mathcal{F}_e$ is the CW likelihood ratio analytically maximized over $h$, $\Phi_0$, $\psi$, and $\iota$. As such, it only depends on $f^{\rm GW}$, $\theta^{\rm GW}$, and $\phi^{\rm GW}$, and gives a good idea of how likely it is to have a CW source at a given frequency and sky location. Figure \ref{fig:f-stat_map} shows $\mathcal{F}_e$ for the dataset with an $A_{\rm CW} = 7.08 \times 10^{-15}$ simulated CW signal used in Section \ref{ssec:study2}. Note how $\mathcal{F}_e$ peaks at the true sky location and frequency of the simulated CW source, and that its spread is relatively small compared to the whole range of parameters. This indicates that by using $\mathcal{F}_e$ as a proposal we can greatly improve convergence, since we will effectively always propose a CW source to the close proximity of the true parameters. This results in at least two orders of magnitude improvement in convergence speed. The maximized values of $h$, $\Phi_0$, $\psi$, and $\iota$ can also be calculated. These can be used to construct informed proposals for these parameters too, however currently we draw those parameters from their prior distributions.

\subsubsection{Trans-dimensional proposals}
\label{ssec:rjmcmc}
The most important proposal in our RJMCMC is the one changing $N_{\rm{CW}}$ or $N_{\rm{GWB}}$. We handle these separately, so we have a proposal which adds or removes a CW signal, and one which adds or removes the GWB from the model. For the GWB adding proposal we draw the GWB amplitude from the prior, while for adding a CW we use $\mathcal{F}_e$ described above in Section \ref{ssec:Fe}. For removing a CW, we randomly select one of the active CW sources to take out from the model.

In practice, these trans-dimensional proposals are much like the within-model ones, with the exception that one should make sure that the proposal densities and priors are properly normalized. While for within-model proposals, an overall normalization of these cancels out in the acceptance probability, having an error in normalizations in a trans-dimensional proposal can lead to biasing our posteriors on $N_{\rm{CW}}$ or $N_{\rm{GWB}}$. To make sure that is not the case, we performed prior recovery tests, i.e.~we set the likelihood to a constant and checked if we recover the priors. \texttt{BayesHopper} passed all such tests.

\subsubsection{Parallel tempering}
Parallel tempering~\cite{PT} (also called Metropolis-coupled MCMC or replica exchange) is commonly used in MCMCs to help explore a multimodal distribution, and it also helps mixing in RJMCMCs (see e.g.~\cite{bayeswave}). We run multiple chains in parallel, with the $i$th chain using a modified likelihood of $p(d|\bm{\theta})^{1/T_i}$, where $T_i$ is the temperature of the $i$th chain. A geometrically spaced temperature ladder is commonly used, where $T_i = c^i$, and $c$ is a constant usually chosen to be between 1.2 and 2. The chains with higher temperature can explore the whole parameter space more easily, and by proposing swaps between chains adjacent on the temperature ladder we help the mixing of the first chain. Eventually, we only record samples from the first chain which has a temperature of 1, because that is the only chain that actually samples from the posterior.

We have found that parallel tempering is essential for exploring different values of $N_{\rm{CW}}$. It results in significantly more $N_{\rm{CW}}$ changing steps in the $T=1$ chain than our dedicated trans-dimensional proposal (see Section \ref{ssec:rjmcmc}). This is especially true for high signal-to-noise ratio (SNR) sources, where it gets hard to let go of a source once the sampler locks onto it. However, a dedicated trans-dimensional step is still needed to effectively explore different $N_{\rm{CW}}$ values at higher temperature chains. This is because parallel tempering can only change $N_{\rm{CW}}$ at a given temperature to its value at an adjacent temperature. E.g.~if a given $N_{\rm{CW}}$ value is not present throughout the temperature ladder, parallel tempering cannot introduce it, but the trans-dimensional proposal can.


\subsubsection{Trans-dimensional post processing for multiple CWs}
\label{ssec:rjmcmc-post}
At the $i$th iteration, the sampler models the dataset with $N_{\rm{CW}, i}$ CWs and labels them CW$_1$, CW$_2$, ..., CW$_{N_{\rm{CW}, i}}$. However, sources with the same label at different iterations do not necessarily correspond to the same physical source, e.g.~CW$_2$ from iteration $i=2$ might not correspond to the same physical source as CW$_2$ from iteration $i=1$, but might correspond to the same physical source as CW$_4$ from iteration $i=5$. To produce Bayes factors and posterior distributions for unique sources, we need a post processing step, which looks through all the samples, identifies individual sources and associates samples with these identified sources. Our post processing procedure works as follows:
\begin{enumerate}
 \item We scan through all the samples and find the one with the highest likelihood ($L_i$). Within the sample with the highest likelihood, we calculate the change in the log likelihood if the given source (labeled by $j$) is turned off ($\Delta \ln L_{i,j}$). The source with the highest $\Delta \ln L_{i,j}$ will be our reference source, denoted with $\bm{\bar{h}}$.
 \item We scan through all the sources in all samples and calculate our distance measure between each source ($\bm{h}$) and the reference source, defined as:
 \begin{equation}
 \mathcal{D}(\bm{\bar{h}}, \bm{h}) = \left(\bm{\bar{h}} - \bm{h}|\bm{\bar{h}} - \bm{h}\right),
 \end{equation}
where $(\bm{a}|\bm{b})$ is the noise-weighted inner product between $\bm{a}$ and $\bm{b}$, defined as:
\begin{equation}
 (\bm{a}|\bm{b}) = \bm{a}^T \bm{\Sigma}^{-1} \bm{b},
\end{equation}
where $\bm{\Sigma}$ is the post-fit noise covariance matrix of the PTA. We then associate sources to the reference source, if $\mathcal{D}(\bm{\bar{h}}, \bm{h})>\hat{\mathcal{D}}$. The optimal value of the $\hat{\mathcal{D}}$ threshold can be determined knowing that \cite{katerina_match_dist}:
\begin{equation}
 \mathcal{D}(\bm{\bar{h}}, \bm{h}) \sim \chi^2_{D-1}, 
\end{equation}

where $D=7$ is the number of signal parameters. To make sure we include all samples corresponding to a given source, we can take the 99.9th percentile of the $\chi^2_{D-1}$ distribution, which sets $\hat{\mathcal{D}} \simeq 25$.
 \item We repeat (i) and (ii) with already assigned sources excluded until all sources are associated with a physical source. We skip any reference source with SNR$<5$, since even if $\bm{h}=0$, $\mathcal{D}(\bm{\bar{h}}, \bm{h}) = (\bm{\bar{h}}|\bm{\bar{h}}) = {\rm SNR}^2$, so with a reference source below an SNR of 5, all low SNR samples would be associated together.
\end{enumerate}

This procedure results in a number of identified physical sources, each with $\bar{n}_{\rm{CW}}$ source samples associated with them. Comparing $\bar{n}_{\rm{CW}}$ to the total number of independent samples, we can calculate Bayes factors with the RJ method (see Section \ref{ssec:bayes-factors}) for each physical source. Analyzing the set of $\bar{n}_{\rm{CW}}$ source samples for each identified source, we can perform source-by-source parameter estimation.

\subsubsection{Calculating Bayes factors}
\label{ssec:bayes-factors}
An RJMCMC algorithm provides a natural way of calculating Bayes factors between different models, by comparing the time the sampler spent at each model. So the Bayes factor between model $\mathcal{A}$ and $\mathcal{B}$ is given by:
\begin{equation}
 {\rm BF}_{\mathcal{A}-\mathcal{B}} = \frac{n_{\mathcal{A}}}{n_{\mathcal{B}}},
\end{equation}
where $n_{\mathcal{A}}$ and $n_{\mathcal{B}}$ are the number of independent samples from model $\mathcal{A}$ and $\mathcal{B}$, respectively. We will refer to this method of calculating Bayes factors as RJ method. It works well if $0.01 \lesssim {\rm BF}_{\mathcal{A}-\mathcal{B}} \lesssim 100$, but otherwise the sampler spends most of the time in one of the models and the estimation becomes unreliable. This can be remedied by the prior re-weighting method, i.e.~by introducing an artificial prior such that $n_{\mathcal{A}} \simeq n_{\mathcal{B}}$. The ideal prior can be determined from a pilot run, and the final result can be corrected for this artificial prior. This method theoretically allows us to calculate Bayes factors of any magnitude, but in practice it only helps a few orders of magnitude, beyond which multiple pilot runs may be needed to find the ideal value of the artificial prior. The statistical uncertainty of ${\rm BF}_{\mathcal{A}-\mathcal{B}}$ can be estimated by assuming that $n_{\mathcal{A}}$ and $n_{\mathcal{B}}$ follow a Poisson distribution. 

The Bayes factors between models $\mathcal{A} : \{N_{\rm{GWB}}=1\}$ and $\mathcal{B} : \{N_{\rm{GWB}}=0\}$, or for $\mathcal{A}:\{N_{\rm{CW}}=1\}$ and $\mathcal{B}:\{N_{\rm{CW}}=0\}$ can also be calculated using the Savage-Dickey density ratio (SD method). The Bayes factor in this case is given by:
\begin{equation}
 {\rm BF}_{\mathcal{A}-\mathcal{B}} = \frac{p(A=0|\mathcal{A})}{p(A=0|d,\mathcal{A})},
\end{equation}
where $A$ is the amplitude parameter of model $\mathcal{A}$, such that $A=0$ is equivalent of model $\mathcal{B}$, $p(A=0|\mathcal{A})$ is the prior at $A=0$, and $p(A=0|d,\mathcal{A})$ is the posterior at $A=0$. In practice, the posterior of $A$ is only available as a histogram from posterior samples, and we approximate $p(A=0|d,\mathcal{A})$ with the lowest amplitude bin\footnote{Note that technically our prior does not include $A=0$ (see Sections \ref{sec:gwb} and \ref{sec:cw}), but the lower limit is typically much smaller than the bin size, so the error introduced by this is much smaller than the effect of having a finite bin size.}. This also means that we have to decide what bin size to use, which will affect our estimation of ${\rm BF}_{\mathcal{A}-\mathcal{B}}$. The smaller the bin size is, the closer we are to truly measuring the posterior at $A=0$, but the fewer independent samples we have in the lowest bin. Another issue arises as ${\rm BF}_{\mathcal{A}-\mathcal{B}}$ gets higher, because $p(A|d,\mathcal{A})$ will peak at a non-zero value of $A$, and it will tail off at low values of $A$ providing very few samples in the lowest bin. A solution to that is doing ``zoom-in'' reruns \cite{katerina_SD}, where we rerun the sampler with a reduced prior range on $A$. The rerun will provide more samples at the lowest bin, and we can apply a correction to the Bayes factor, because we know the fraction of samples falling into the reduced prior range from the first run. This procedure can be repeated multiple times, until we get a sufficient number of samples at the lowest bin. The statistical uncertainty on the SD Bayes factor can be estimated by assuming that the number of independent samples in the lowest bin follows a Poisson distribution.


\section{Results}
\label{sec:results}
We tested \texttt{BayesHopper} on different simulated datasets to demonstrate its capabilities. All datasets correspond to a simulated pulsar timing array with 20 pulsars  at random sky locations (see red crosses on Figure \ref{fig:f-stat_map}) and the same white nose level of $\sigma = 0.5 \ \mu$s for each pulsar. Our datasets does not include RN. All pulsars of the simulated dataset were observed for 10 years, evenly spaced at every 30 days. We added different signals to this dataset to test different parts of our algorithm. First, we looked at datasets with only a GWB (see Section \ref{ssec:study1}) or only a single CW in them (see Section \ref{ssec:study2}). Our results for a dataset with three CWs is presented in Section \ref{ssec:study2b}. Finally, Section \ref{ssec:study4} describes our analysis of a dataset with both a GWB and three CWs in it.

\subsection{Stochastic background}
\label{ssec:study1}

We simulated a GWB with different amplitudes resulting in 7 datasets. We ran \texttt{BayesHopper} on these datasets in GWB-only mode, i.e.~allowing $N_{\rm{GWB}}$ to be either 1 or 0, but requiring $N_{\rm{CW}}=0$. We then calculated the Bayes factor between the $N_{\rm{GWB}}=1$ and $N_{\rm{GWB}}=0$ models, denoted as ${\rm BF}_{\rm GWB}$, using both RJ and SD methods described in Section \ref{ssec:bayes-factors}.

\begin{figure}[htb]
 \centering
    \begin{subfigure}[b]{0.625\textwidth}
            \includegraphics[width=\linewidth]{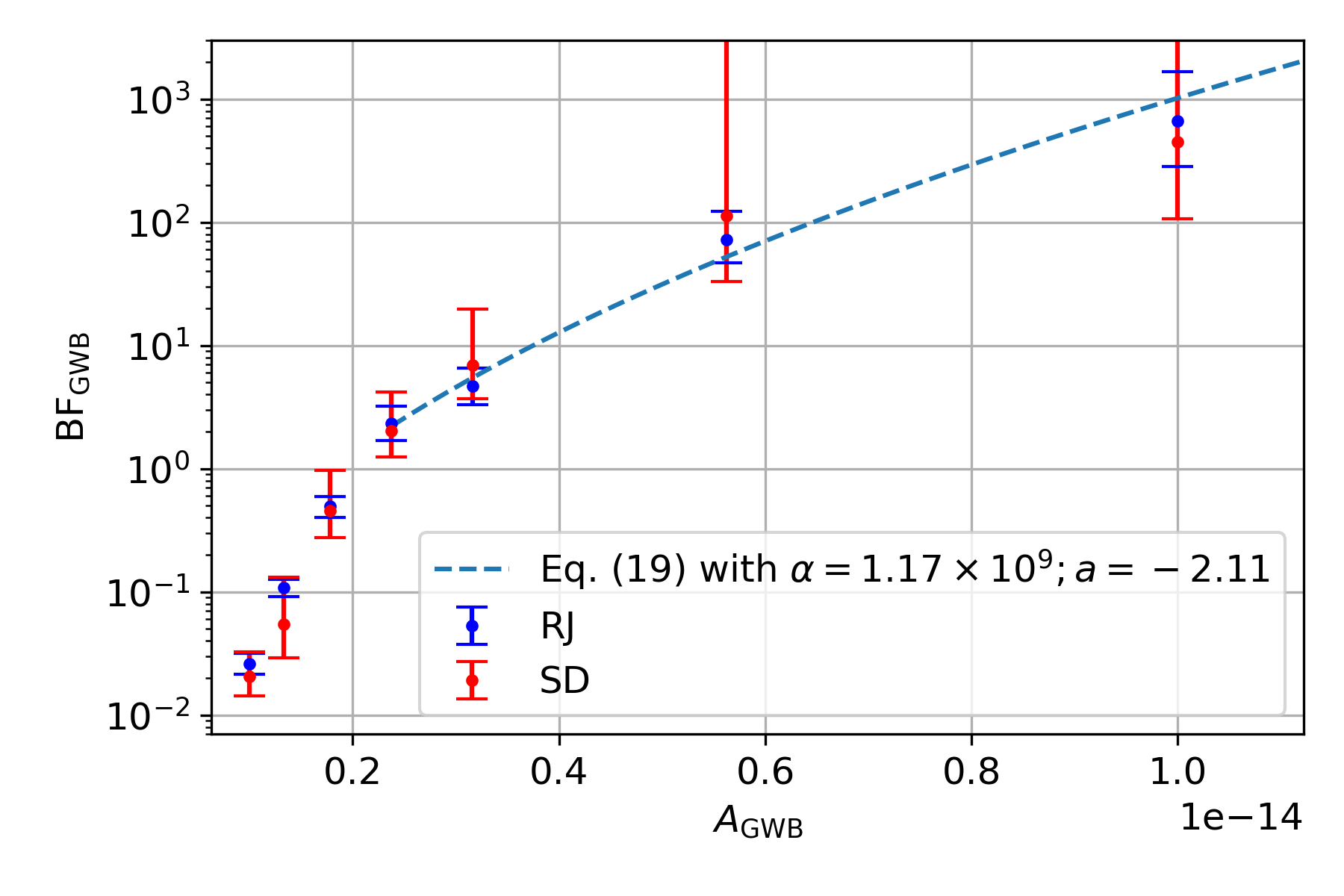}
            \caption{}
            \label{fig:study1BFs}
    \end{subfigure}%
    \begin{subfigure}[b]{0.375\textwidth}
            \includegraphics[width=\linewidth]{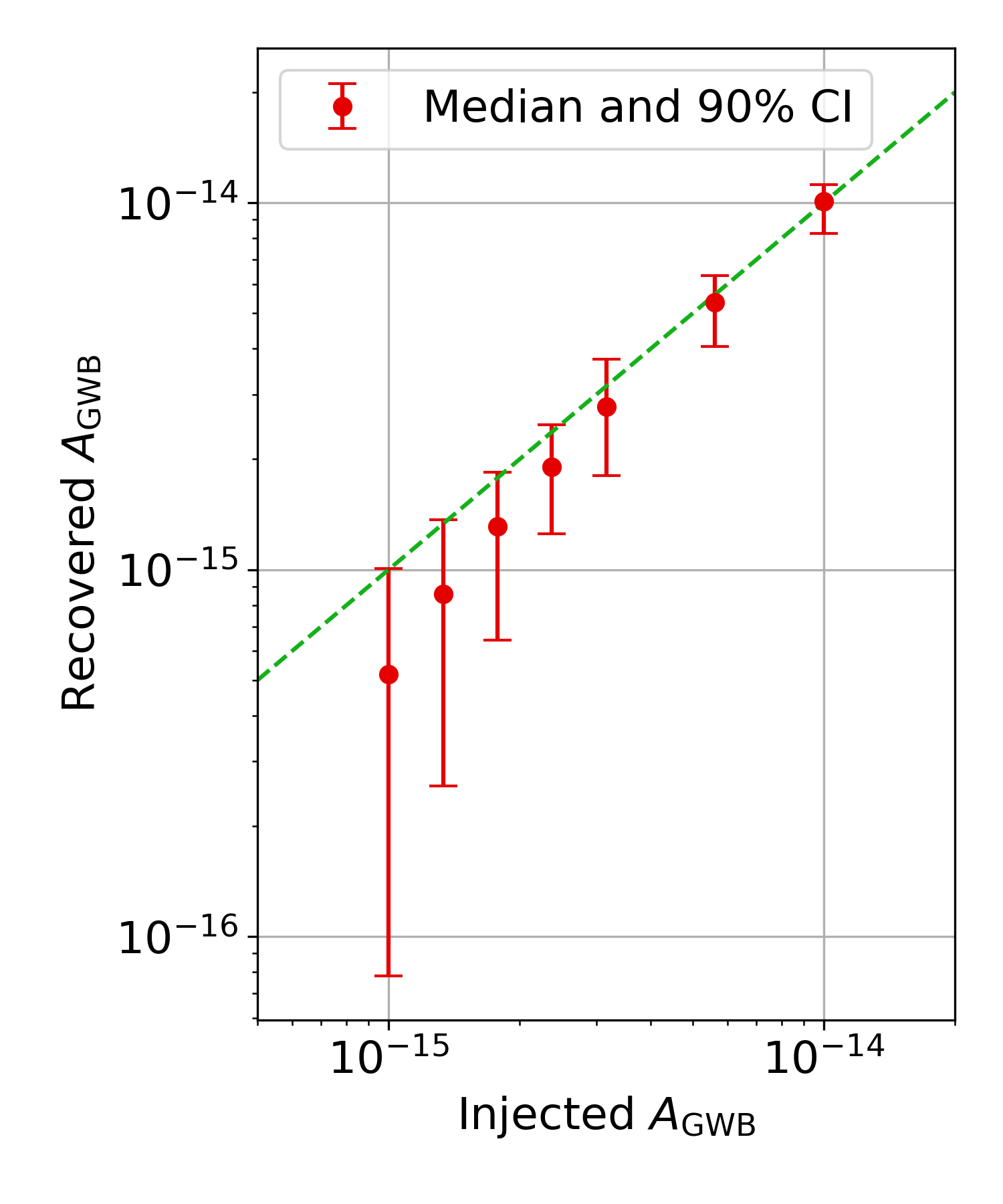}
            \caption{}
            \label{fig:study1_amp_rec}
    \end{subfigure}%
 \caption{(a) Bayes factors for a GWB signal for different levels of the background. Bayes factors obtained using Savage-Dickey density ratio are labeled with red, while Bayes factors calculated using RJMCMC methods are blue. Note that the two methods give consistent results. Error bars indicate 90\% credible intervals obtained by methods described in Section \ref{ssec:bayes-factors}. The dashed line shows a fit in the high SNR regime using the theoretical scaling described by (\ref{eq:gwb_bf_theory}). (b) Recovered values of $A_{\rm GWB}$ as a function of the injected values present in the dataset. Medians and 90\% credible intervals are shown for each dataset. The dashed line indicates perfect recovery, i.e.~where the injected and recovered amplitudes are the same.}
 \label{fig:study1}
\end{figure}

Figure \ref{fig:study1BFs} shows ${\rm BF}_{\rm GWB}$ as a function of the amplitude of the simulated GWB ($A_{\rm GWB}$). Bayes factors obtained using Savage-Dickey density ratio are labeled with red, while Bayes factors calculated using RJMCMC methods are blue. Error bars indicate 90\% credible intervals obtained by methods described in Section \ref{ssec:bayes-factors}. Note that ${\rm BF}_{\rm GWB}$ calculated with the RJ method, our primary method of calculating Bayes factors in an RJMCMC, are consistent with the ones calculated using the SD method, which is widely used in analysis of pulsar timing arrays.

For the two highest amplitude GWBs, our ${\rm BF}_{\rm GWB}$ calculated using the SD method is consistent with it being infinity. This is due to the fact that for high-amplitude signals, we get an increasingly small amount of samples at the lowest $A_{\rm GWB}$ bin, which is where the SD method estimates the Bayes factor. In this case, even after applying the ``zoom-in'' procedure described in Section \ref{ssec:bayes-factors}, we still only get a few samples in the lowest amplitude bin, which is consistent with being zero at the 90\% credibility level, so the resulting Bayes factor is consistent with infinity. One can do multiple ``zoom-in'' reruns to remedy this, but we decided not to carry out those, since our SD Bayes factors are just a secondary product meant to be a sanity check of our RJ Bayes factors.

To check if the ${\rm BF}_{\rm GWB}$ values we get are consistent with what we would expect for a GWB observed by a pulsar timing array, we derive the theoretically expected scaling of ${\rm BF}_{\rm GWB}$ with $A_{\rm GWB}$, and compare it to our results.  Using the Laplace approximation the Bayes factor can be written as:
\begin{equation}
 \ln {\rm BF} = \ln \mathcal{L_{\rm max}} + \ln \left[ p(\theta_{\rm ML}) V_{\rm post} \right],
\label{eq:bf-general}
\end{equation}
where $\mathcal{L_{\rm max}}$ is the maximum likelihood ratio, and the second term is the Occam penalty, where $p(\theta_{\rm ML})$ is the prior at the maximum likelihood parameters, and $V_{\rm post}$ is the posterior volume. Note that if $p(\theta)$ is uniform in all parameters, then the second term in (\ref{eq:bf-general}) is just the logarithm of the ratio between posterior and prior volumes. 

For the GWB model, the maximum likelihood ratio is given by $\ln \mathcal{L_{\rm max}} = \rho^2/2$, where $\rho$ is the optimal statistic SNR \cite{optimal_stat}. The scaling of $\rho^2$ with $A_{\rm GWB}$ is given by \cite{xavi_scaling}:
\begin{equation}
 \rho^2 (A_{\rm GWB}; \alpha) = \alpha \left[{}_2 F_1 \left(2, \gamma^{-1}; 1+\gamma^{-1}; -\frac{12 \pi^2 f^{\gamma} P_{\rm WN}}{A_{\rm GWB}^2 f_{\rm yr}^{4/3}} \right) f \right]_{f=f_{\rm min}}^{f_{\rm max}},
\label{eq:hypergeom}
\end{equation}
where $\alpha$ is a constant independent of $A_{\rm GWB}$, ${}_2 F_1$ is the ordinary hypergeometric function, $\gamma=13/3$ is the spectral slope of the GWB, and $P_{\rm WN} = 2 \sigma^2 \Delta t$ is the PSD of the white noise, where $\Delta t = 30$ days is cadence. $f_{\rm min} = 1/T_{\rm obs}$ is the minimum, while $f_{\rm max} = 30/T_{\rm obs}$ is the maximum frequency used in our GWB model (see Section \ref{sec:gwb}). In case of a fixed spectral index GWB model, we only have one parameter, so in the high SNR regime, the Occam penalty terms will not depend strongly on $A_{\rm GWB}$, so we will assume they are constant. So we expect the Bayes factor to be given by:
\begin{equation}
\ln {\rm BF}_{\rm GWB} \approx \frac{\rho^2(A_{\rm GWB}; \alpha)}{2} + a,
\label{eq:gwb_bf_theory}
\end{equation}
where $a$ is a free parameter describing the $A_{\rm GWB}$-independent contribution from the Occam terms. Note that this expression has two free parameters, $a$ and $\alpha$ from (\ref{eq:hypergeom}).

In Figure \ref{fig:study1BFs} we fit the highest 4 ${\rm BF}_{\rm GWB}$ values from the RJ method with this expected relation. We can see that (\ref{eq:gwb_bf_theory}) fits the results well, indicating that the Bayes factors we get are consistent with the theoretical expectations.

Figure \ref{fig:study1_amp_rec} shows the median recovered $A_{\rm GWB}$ values with their 90\% credible intervals as a function of the injected amplitude value present in the dataset. We can see that as $A_{\rm GWB}$ increases the recovered value approaches the true value, as expected when we transition from an upper-limit type of result to a detection. We can also see how the relative uncertainty on $A_{\rm GWB}$ decreases as the injected amplitude (and thus the SNR) increases.

\subsection{Single continuous wave}
\label{ssec:study2}

We added a single CW signal to our dataset with different amplitudes ($A_{\rm CW}$), resulting in 8 different datasets. Other parameters of the source were kept fixed throughout these datasets, and were the same as for source \#1 in Table \ref{tab:injected_cws}. We ran \texttt{BayesHopper} on these datasets in CW-only mode with $N^{\rm{max}}_{\rm{CW}} = 1$, i.e.~we fixed $N_{\rm{GWB}}$ to be zero, and allowed $N_{\rm{CW}}$ to be either 0 or 1. Then, we calculated Bayes factors between the $N_{\rm{CW}}=1$ and $N_{\rm{CW}}=0$ models, denoted as ${\rm BF}_{\rm CW}$, with both RJ and SD methods described in Section \ref{ssec:bayes-factors}.

\begin{figure}[htb]
 \centering
       \begin{subfigure}[b]{0.625\textwidth}
            \includegraphics[width=\linewidth]{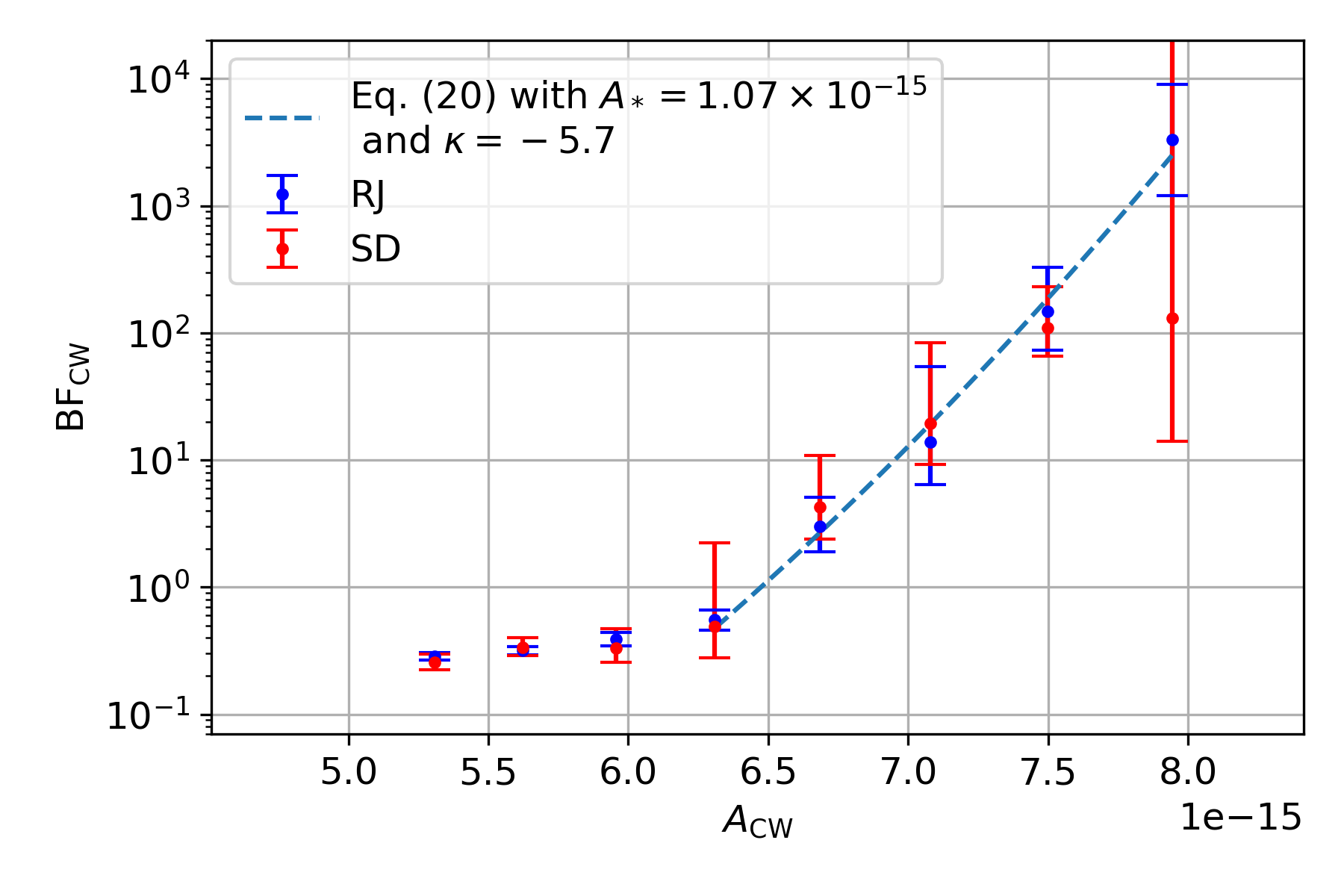}
            \caption{}
            \label{fig:study2BFs}
    \end{subfigure}%
    \begin{subfigure}[b]{0.375\textwidth}
            \includegraphics[width=\linewidth]{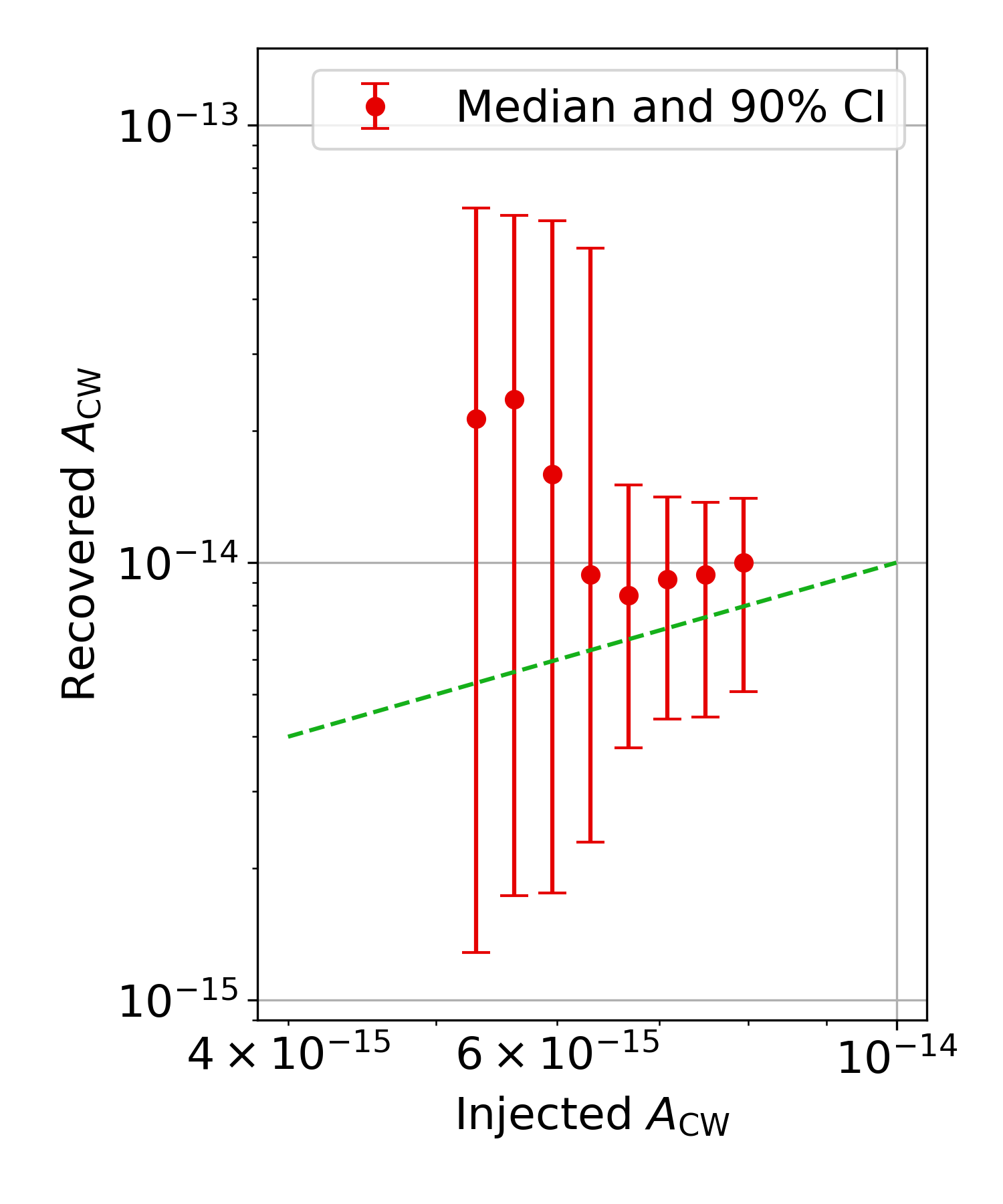}
            \caption{}
            \label{fig:study2_amp_rec}
    \end{subfigure}%
 \caption{(a) Bayes factors of a CW signal for different injected amplitudes. Bayes factors obtained using Savage-Dickey density ratio are labeled with red, while Bayes factors calculated using RJMCMC methods are blue. Note that the two methods give consistent results. Error bars indicate 90\% credible intervals obtained by methods described in Section \ref{ssec:bayes-factors}. The dashed line shows a fit in the high SNR regime using the theoretical scaling of ${\rm BF}_{\rm CW}$ from (\ref{eq:cw_bf_theory}), where we find $A_*= 1.07 \times 10^{-15}$ and $\kappa = -5.7$. The good fit indicates that the Bayes factors we get are consistent with the theoretical scaling. (b) Recovered values of $A_{\rm CW}$ as a function of the injected values present in the dataset. Medians and 90\% credible intervals are shown for each dataset. The dashed line indicates perfect recovery, i.e.~where the injected and recovered amplitudes are the same.}
 \label{fig:study2}
\end{figure}

Figure \ref{fig:study2BFs} shows ${\rm BF}_{\rm CW}$ as a function of $A_{\rm CW}$, both for RJ (blue) and SD (red) methods. Error bars indicate 90\% credible intervals obtained by methods described in Section \ref{ssec:bayes-factors}. Note that ${\rm BF}_{\rm CW}$ calculated with the RJ method, our primary method of calculating Bayes factors in an RJMCMC, are consistent with the ones calculated using the SD method, which is widely used in data analysis of pulsar timing arrays.

To check if the ${\rm BF}_{\rm CW}$ values we get are consistent with what we would expect for a CW signal observed by a pulsar timing array, in the following we derive the theoretically expected scaling of ${\rm BF}_{\rm CW}$ with $A_{\rm CW}$, and compare it to our results. The general expression of a Bayes factor is given by (\ref{eq:bf-general}). For the CW case, $\ln  \mathcal{L_{\rm max}} = {\rm SNR}^2/2$. For slowly varying priors, and large SNR,  the posterior volume can be approximated as $V_{\rm post} \approx (\det (\Gamma/2\pi))^{-1/2}$, where $\det (\Gamma)$ is the determinant of the Fisher information matrix, which scales with the SNR as $\det(\Gamma) \propto {\rm SNR}^{2 D}$, where $D=7$ is the model dimension. Since ${\rm SNR} \propto A_{\rm CW}$, we get the following theoretical scaling of the CW Bayes factor:
\begin{equation}
 \ln {\rm BF}_{\rm CW} \approx \frac{1}{2} \left(\frac{A_{\rm CW}}{A_*}\right)^2 - D \ln\left(\frac{A_{\rm CW}}{A_*}\right)+ \kappa
 \label{eq:cw_bf_theory},
\end{equation}
where $A_*$ fixes the scaling between amplitude and SNR and $\kappa$ is a constant.

Figure \ref{fig:study2BFs} shows the fit to ${\rm BF}_{\rm CW}$ using the scaling described by (\ref{eq:cw_bf_theory}). We can see that the scaling fits well in the high SNR regime, indicating that our Bayes factors are consistent with the theoretical expectation.

Figure \ref{fig:study2_amp_rec} shows the median recovered $A_{\rm CW}$ values with their 90\% credible intervals as a function of the injected amplitude value present in the dataset. We can see that as $A_{\rm CW}$ increases the recovered value approaches the true value, as expected when we transition from an upper-limit type of result to a detection. We can also see how the relative uncertainty on $A_{\rm CW}$ decreases as the injected amplitude (and thus the SNR) increases.

\subsection{Multiple continuous waves}
\label{ssec:study2b}

In this section we analyzed a single dataset, where we added 3 different CW sources. Parameters of the three source are given in Table \ref{tab:injected_cws}. We analyzed this dataset by \texttt{BayesHopper} in the CW-only mode, with $N^{\rm{max}}_{\rm{CW}} = 10$, i.e.~we fixed $N_{\rm{GWB}}$ to be zero, and allowed $N_{\rm{CW}}$ to vary between 0  and 10, with a uniform prior on $N_{\rm{CW}}$.

\begin{table}[htbp]
\caption{\label{tab:injected_cws}Parameters of simulated CW signals.}
\footnotesize
\begin{tabular}{@{}lllllllll}
\br
Source number&$f^{\rm GW}$ [nHz]&$A_{\rm CW}$ [10$^{-15}$]&$\theta^{\rm GW}$&$\phi^{\rm GW}$&$\Phi_0$&$\psi$&$\iota$&SNR\\
\mr
\#1&8&7.5&$\pi/3$&4.5&1.0&1.0&1.0&8.6\\
\#2&20&16.4&$2\pi/3$&1.0&0.5&0.0&0.8&7.0\\
\#3&60&34.1&$\pi/6$&2.0&0.0&1.2&0.3&6.9\\
\br
\end{tabular}\\

\end{table}
\normalsize

\begin{figure}[htb]
 \centering
   \includegraphics[width=0.9\textwidth]{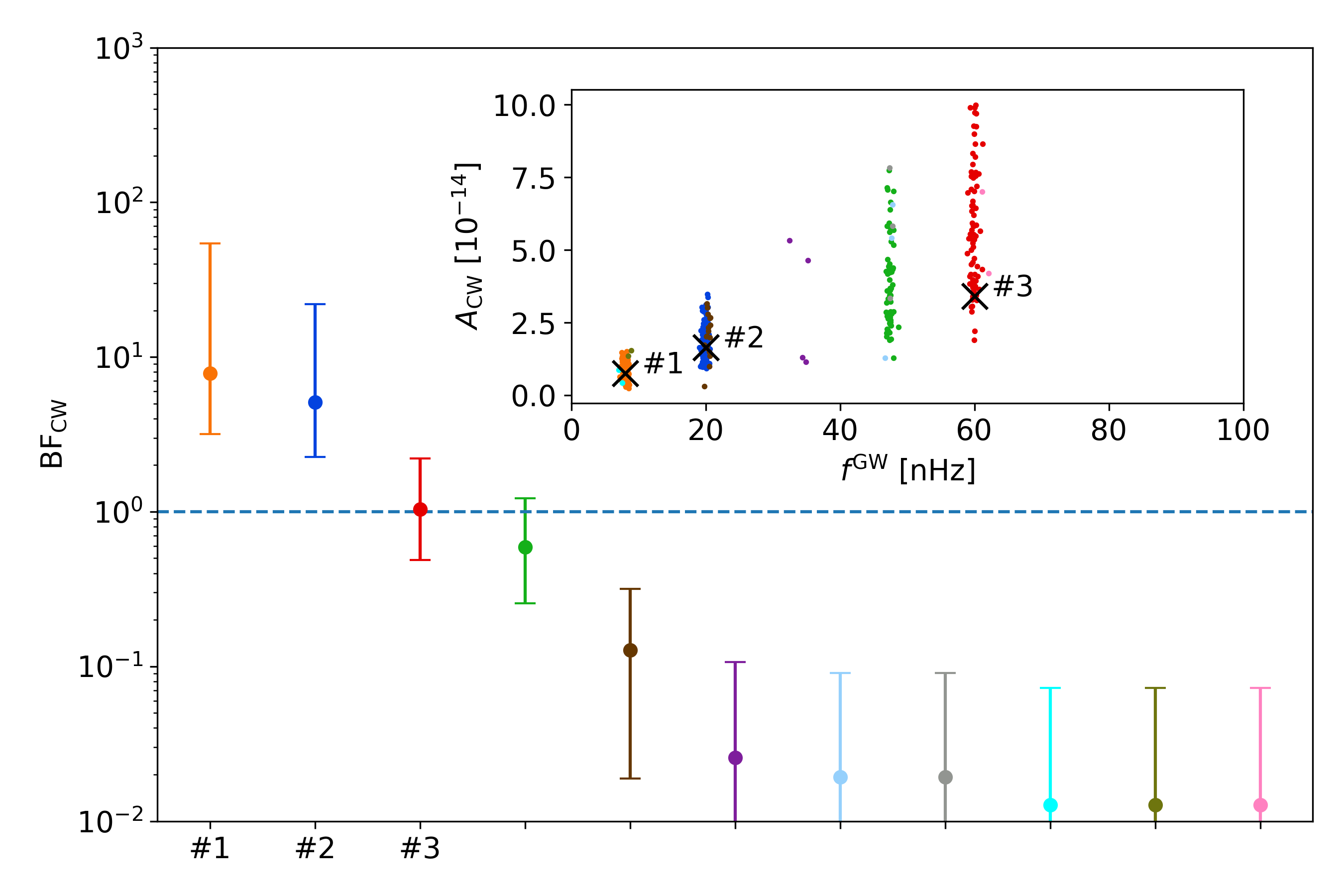}
 \caption{Bayes factors of different CW sources found by the source identification post processing step. Inset shows the distribution of $f_{\rm GW}$ and $A_{\rm CW}$ for each source, with true parameter values of the simulated sources indicated by $\times$ symbols. Note that the three simulated CWs (see Table \ref{tab:injected_cws}) were clearly found with ${\rm BF}_{\rm CW}$ increasing with simulated SNRs.}
 \label{fig:study2bBFs}
\end{figure}

We used the post processing procedure described in Section \ref{ssec:rjmcmc-post} to identify physical CW sources, and calculated the Bayes factors for each of those. Figure \ref{fig:study2bBFs} shows ${\rm BF}_{\rm CW}$ for each identified source, with the inset showing $f_{\rm GW}-A_{\rm CW}$ posterior samples for each source. We  can see that the source separation algorithm does a fair job separating the sources, and that the posterior coincides with the true values of $f_{\rm GW}$ and $A_{\rm CW}$ marked with $\times$ symbols. Note that the three most significant sources identified correspond to the three simulated CW signals added to the dataset. As expected, the ${\rm BF}_{\rm CW}$ values we get for these increase with the SNR of the simulated signal. The most significant source not associated with an injected source has a Bayes factor slightly below 1. We re-analyzed the dataset with a different realization of the WN, and found that this source disappeared, but another source with similar significance appeared. This suggests that this source corresponds to the sampler trying to fit some noise fluctuation with the CW model, and it is reassuring to see that we slightly disfavor it being a real CW signal.

\subsection{Stochastic background and continuous waves}
\label{ssec:study4}
To assess how the GWB model in itself affects the CW model, we first analyzed the same dataset as in Section \ref{ssec:study2b}, but allowing the sampler to also vary $N_{\rm{GWB}}$, instead of setting it to zero. We have found that the three simulated signals have ${\rm BF}_{\rm CW}$ values consistent with the CW-only run. So in this particular case, including a GWB signal in the model does not seem to affect our results for the CWs. This is not surprising, since we only have three CW sources, all of which are relatively strong. We expect the effect of the GWB model soaking up CW sources to play a stronger role when many low-SNR CWs are present. 

\begin{figure}[htb]
 \centering
   \includegraphics[width=0.9\textwidth]{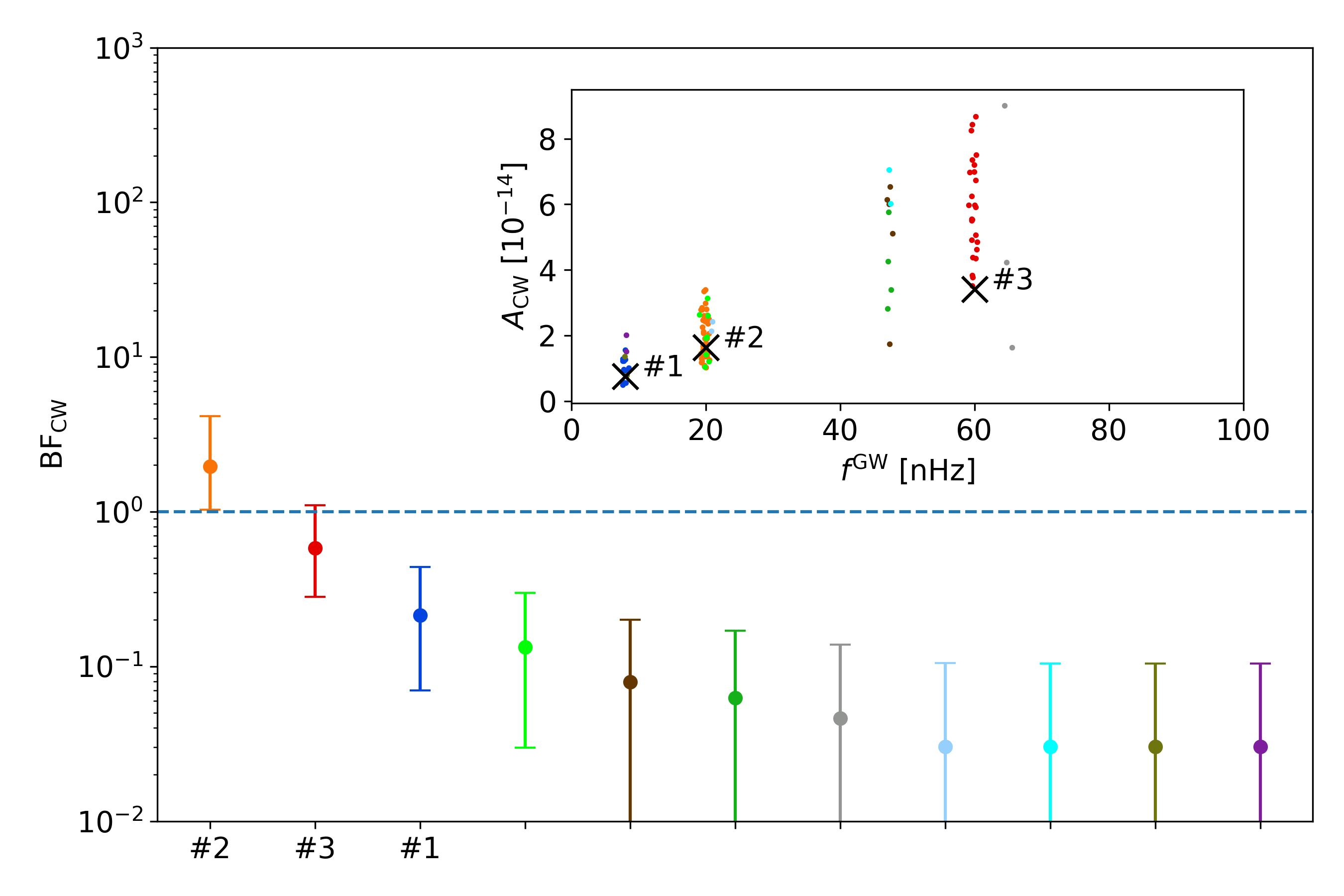}
 \caption{Bayes factors of different physical CW sources found by the source identification post processing step (see Section \ref{ssec:rjmcmc-post}). Inset shows the distribution of $f_{\rm GW}$ and $A_{\rm CW}$ for each source, with true parameter values of the simulated sources indicated by $\times$ symbols. This dataset includes a simulated GWB with $A_{\rm GWB} = 3.98 \times 10^{-15}$ and CW sources described in Table \ref{tab:injected_cws}. Note that only the moderate-SNR source at moderate frequency remained significant, while both the high-SNR source at low frequency and the moderate-SNR source at high frequency is unfavored when a GWB of this level is present.}
 \label{fig:study4_highgwb_BFs}
\end{figure}

To test BayesHopper on a dataset with both CWs and a GWB in it, we used the sources described in Table \ref{tab:injected_cws} and we also added an isotropic GWB signal with $A_{\rm GWB} = 3.98 \times 10^{-15}$. The GWB amplitude is near the threshold level identified in Figure \ref{fig:study1BFs}, and in the combined search yielded a Bayes factor of ${\rm BF}_{\rm GWB}=5.18^{+4.82}_{-1.99}$. Figure \ref{fig:study4_highgwb_BFs} shows the identified sources with their ${\rm BF}_{\rm CW}$ values and $f_{\rm GW}-A_{\rm CW}$ posterior samples. The three most significant sources are still the three simulated CWs, similarly to what we have seen in Section \ref{ssec:study2b}, but their significance changed considerably. While the Bayes factor of source \#2 stayed about the same, source \#1 and \#3 show a significant decrease in their Bayes factors. For source \#1, this can be explained by the increased level of the noise at low frequencies due to the GWB. This explanation is also consistent with the fact that the $A_{\rm CW}$ posterior extends to much lower values for the dataset without GWB (compare inset figures of Figure \ref{fig:study2bBFs} and \ref{fig:study4_highgwb_BFs}). Source \#3 is at much higher frequency, where the increase in the noise level due to the GWB is negligible. However, it was already a low-significance CW source, so in this case we think that the decrease in ${\rm BF}_{\rm CW}$ can be explained by the fact that the GWB model might be able to partially fit the contribution of source \#3, which might be statistically preferred over the more complex CW model.

\begin{figure}[htb]
 \centering
   \includegraphics[width=0.9\textwidth]{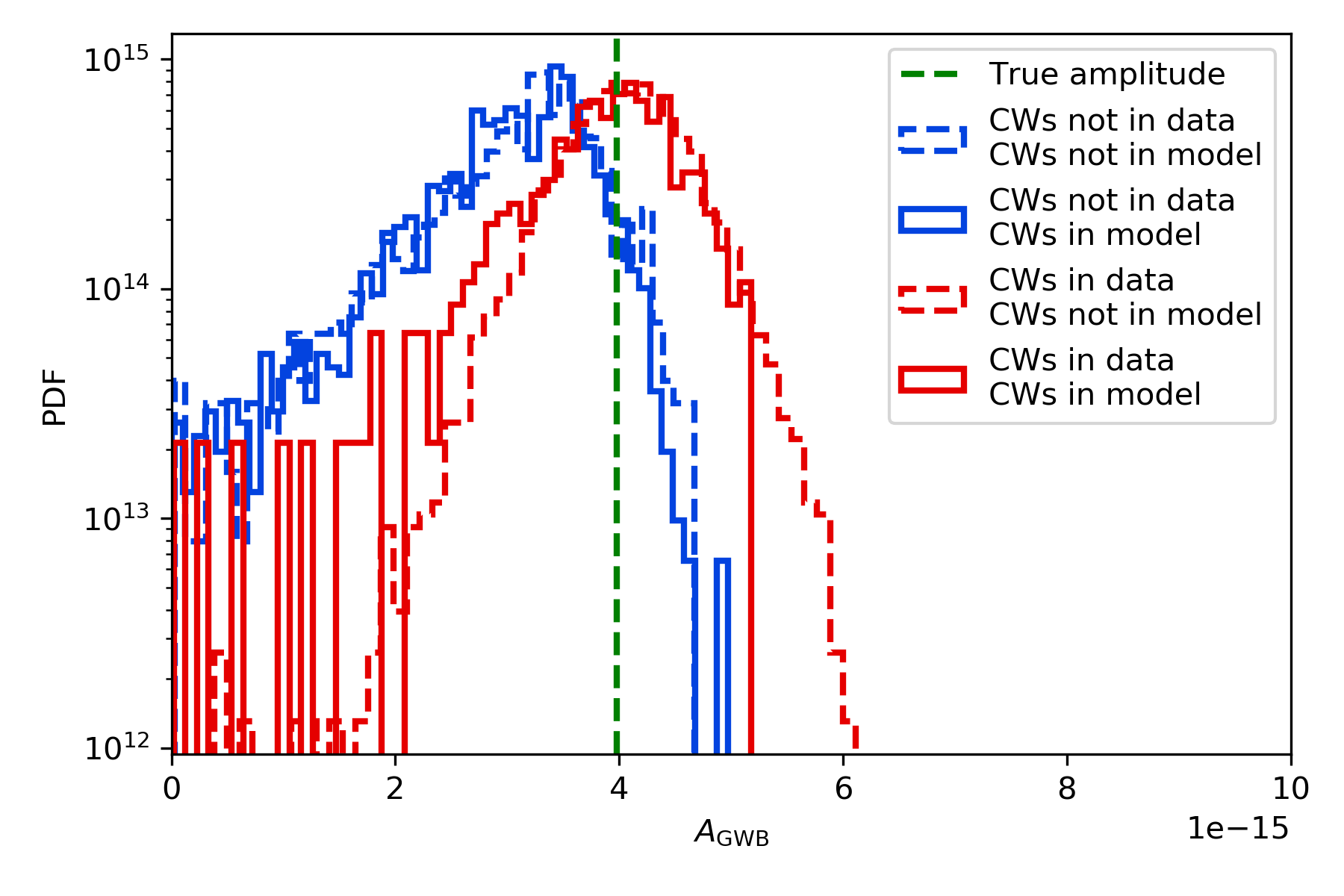}
 \caption{Posterior distribution of $A_{\rm GWB}$ for four different runs. Each dataset included a stochastic background with $A_{\rm GWB} = 3.98 \times 10^{-15}$. Blue histograms show results for a dataset with only this GWB signal, while red histograms are for a dataset that also included the 3 CW signals from Table \ref{tab:injected_cws}. Solid lines show results for runs where we allowed the sampler to include CWs in the model, while dashed lines represent runs where we did not allow the sampler to include CWs in the model. We see that the GWB model absorbs some power from the CWs. The ${\rm BF}_{\rm GWB}$ values obtained for these runs are reported in Table \ref{tab:Bayes-factors}}
 \label{fig:study4_highgwb_gwb_amp}
\end{figure}

We have seen that the significance of CW sources can be affected by the presence of a GWB. To investigate whether CWs can affect the significance of GWBs, we analyze 2 datasets: one with no CWs and a GWB with $A_{\rm GWB} = 3.98 \times 10^{-15}$, and one with the same GWB and the three CWs described in Table \ref{tab:injected_cws}. We analyze both of these datasets with two settings: one where our model can only contain a GWB signal, and one where it can contain a GWB signal and up to 10 CW signals. Table \ref{tab:Bayes-factors} shows the ${\rm BF}_{\rm GWB}$ values we get for each of these runs. We can see that when no CWs are present in the dataset, it does not matter if we include CWs in the model or not, we get ${\rm BF}_{\rm GWB} \sim 1$ in both cases. If CWs are present in the dataset and we include them in our model, we get a slightly higher ${\rm BF}_{\rm GWB} \sim 4$. This is consistent with the fact that ${\rm BF}_{\rm CW}$ decreased when we included GWBs, because part of the CW signals can be fit with the GWB model. Finally, if CWs are present in the dataset, but we do not model them, we see an even higher increase in the GWB significance (${\rm BF}_{\rm GWB} \sim 120$) indicating that in this case the GWB model tries to fit as much of the CW signals as it can. Figure \ref{fig:study4_highgwb_gwb_amp} shows posterior distributions of $A_{\rm GWB}$ for the same runs. We see again that the GWB model takes some power from the CWs, increasing the significance of the GWB, and decreasing the significance of the CWs. Note that the change in the recovered amplitude is consistent with the bias we see in Figure \ref{fig:study1_amp_rec}.

\begin{table}[htbp]
\caption{\label{tab:Bayes-factors}Dependence of ${\rm BF}_{\rm GWB}$ on CWs. ${\rm BF}_{\rm GWB}$ values are shown for four different runs. Each dataset included a stochastic background with $A_{\rm GWB} = 3.98 \times 10^{-15}$, and datasets reported in the second column also included the 3 CW signals from Table \ref{tab:injected_cws}. All runs included a GWB in the model, and runs reported in the second row also allowed for CW signals.}
\footnotesize
\begin{tabular}{@{}l|ll}
\br
 &CWs not in data&CWs in data\\
\mr
CWs not in model & $1.018^{+0.067}_{-0.063}$ & $124^{+35}_{-23}$ \\
CWs in model & $1.00^{+0.14}_{-0.12}$ & $3.9^{+3.6}_{-1.5}$ \\ 
\br
\end{tabular}\\

\end{table}
\normalsize

\section{Conclusion}
\label{sec:conclusion}

We have presented \texttt{BayesHopper}, an RJMCMC algorithm for jointly searching for isolated CW sources and an isotropic GWB in PTA data. We have tested it on various simulated datasets, and have found that for the CW-only and GWB-only datasets it delivers results consistent with methods currently used in the PTA community. For datasets with multiple CWs in them, we have illustrated that an additional post processing step is needed to identify physical CW sources from the MCMC samples, and introduced a method which can separate these sources by finding the maximum likelihood parameters of sources and comparing it to other source parameters in the samples. For the full problem of analyzing a dataset with both a GWB and multiple CWs in it, we have seen some examples of how the CW and GWB models interact. While we have seen that allowing to fit for a GWB when no GWB is present does not change the CW results (at least in the one case we looked into), we have seen that the significance of CW sources can decrease in presence of a GWB because of two different reasons: i) the GWB increases the noise level, which can make a CW source at low frequency insignificant; ii) sometimes the GWB model can partially fit the CW signal, and having only 1 parameter, it can be preferred by the Bayesian analysis' natural parsimony, even if it provides a worse fit than the CW model, especially if the CW source had low-SNR to start with. This interaction between the GWB and CW models is what makes such a joint analysis interesting and worthwhile, and we plan to investigate such interactions in greater depth in upcoming studies.

Beyond being the first implementation of a joint analysis of CWs and a GWB in PTAs, \texttt{BayesHopper} is also the first algorithm which can calculate Bayes factors of a CW or a GWB based on an RJMCMC, instead of the Savage-Dickey density ratio method. It would be beneficial to run \texttt{BayesHopper} on real PTA data, to provide an independent measurement of the Bayes factors.

Work is currently underway in testing \texttt{BayesHopper} on a realistic cosmological simulation \cite{LukeSMBHCatalog}, where the GWB is not manually added as an isotropic background, but instead builds up from simulated BBHs in the Universe. This will provide a more realistic test of the algorithm, since the astrophysical background is expected to be built from a collection of individually unresolvable CWs. Indeed, the ``true'' model of the PTA dataset would be just a large collection of CWs, but due to the parsimony of Bayesian inference, a model with fewer parameters including a few CWs and an isotropic GWB will be preferred (see e.g.~\cite{when_stoch}). Such a study on realistic simulations will prepare us for analyzing real PTA datasets with our full model including multiple CWs and a GWB. Such an analysis of the NANOGrav 12.5-year dataset is planned for the near future.

We are also planning to build on the RJMCMC engine developed for \texttt{BayesHopper} to create a PTA version of the BayesWave \cite{bayeswave} algorithm, which is used to search for unmodeled transient GWs in the data of ground-based interferometric GW detectors. This would only require changing the sinusoidals in our CW model to Morlet-Gabor wavelets used in BayesWave, and we would be in a position to search for unmodeled transients in PTA data.

\ack
We acknowledge the help of Logan O'Beirne in installing and using the \texttt{Enterprise} software package.
We appreciate the support of the NSF Physics Frontiers Center Award PFC-1430284.
Computations were performed on the Hyalite High Performance Computing System, operated and supported by University Information Technology Research Cyberinfrastructure at Montana State University.

\section*{References}
\bibliography{bayeshopper_bib}{}
\bibliographystyle{unsrt_et_al}

\end{document}